\documentclass[a4paper,pre,aps,preprint]{revtex4}

\usepackage{ulem}
\usepackage{amsfonts}
\usepackage{amsmath}
\usepackage{amssymb}
\usepackage{graphicx}
\usepackage{mathrsfs}
\usepackage[font={small,it}]{caption}

\newcommand{\ab}{\alpha\beta}
\newcommand{\partialP}{\frac{\partial}{\partial P}}
\newcommand{\partialR}{\frac{\partial}{\partial R}}
\newcommand{\half}{\frac{1}{2}}

\newcommand{\aabb}{\alpha\alpha',\beta\beta'}

\newcommand{\sign}{\mathop{\mathrm{sign}}\nolimits}

\begin{document}


\title{On Filtering Schemes in 
the Quantum-Classical Liouville Approach
to Non-adiabatic Dynamics}

\author{Daniel Uken}
\affiliation{School of Chemistry and Physics, University of KwaZulu-Natal\\
Private Bag X01, Scottsville 3209 Pietermaritzburg, South Africa}

\author{Alessandro Sergi} 
\affiliation{School of Chemistry and Physics, University of KwaZulu-Natal\\
Private Bag X01, Scottsville 3209 Pietermaritzburg
and National Institute for Theoretical Physics (NITheP), KwaZulu-Natal, South Africa}

\author{Francesco Petruccione}
\affiliation{School of Chemistry and Physics
and National Institute for Theoretical Physics,
University of KwaZulu-Natal, Westville Campus, Private Bag X54001,
Durban 4000, South Africa 
and National Institute for Theoretical Physics (NITheP), KwaZulu-Natal, South Africa}

\begin{abstract}
We study a number of filtering schemes for the reduction of the statistical error
in non-adiabatic calculations by means of the quantum-classical Liouville equation.
In particular, we focus on a scheme based on setting a threshold value
on the sampling weights, so that when the threshold is overcome
the value of the weight is reset, and on another approach which prunes
the ensemble of the allowed non-adiabatic transitions
according to a generalised sampling probability.
Both methods have advantages and drawbacks, however their combination
drastically improves the performance of 
an algorithm known as the Sequential Short Time Step Propagation
[D. MacKernan et al., J. Phys: Condens. Matter {\bf 14} 9069 (2002)],
which is derived from a simple first order expansion of the quantum-classical propagator.
Such an algorithm together with the combined filtering procedures produce
results that compare very well with those obtained
by means of numerically ``exact'' quantum calculations for the spin-boson model,
even for intermediate and strong coupling regimes.
\end{abstract}


\maketitle

\section{Introduction}

In the field of condensed matter, many systems can be modeled using a quantum
subsystem coupled to a classical bath. When energy is free
to be exchanged between the subsystem and the bath, the resulting dynamics is
known to be non-adiabatic.
This type of dynamics is very difficult
to simulate due to the quantum back-reaction of the subsystem onto the 
bath~\cite{aleksandrov,gerasimenko,boucher,zhang,balescu}.
A number of numerical methods have been proposed for the calculation of non-adiabatic
dynamics on a computer, the most common of which are so-called surface hopping 
schemes~\cite{tully,miller,pechukas,pechukas2,heller,shenvi,berne}.
More recently, an approach
based on the quantum-classical Liouville equation has been applied
with success to condensed matter systems~\cite{qcl5,qcl6,qcl8,qcl9,ilya,hanna,geva}.
This approach
allows one to construct a proper formulation of the statistical mechanics of 
quantum-classical systems~\cite{nielsen,b-silurante}
which can also be generalised to situations where the bath follows
a non-Hamiltonian dynamics~\cite{sergi-temp1,sergi-temp2}.

A simple and efficient algorithm suited for the computer simulation of the
quantum-classical Liouville equation is the 
Sequential Short-Time Propagation (SSTP) algorithm~\cite{sstp}.
The SSTP algorithm is based on a first order expansion in time of
the Dyson form of the quantum-classical propagator and, when combined with
the momentum-jump approximation~\cite{momj,momj2}, leads to a representation of 
non-adiabatic dynamics in terms of piecewise adiabatic
trajectories of the bath coordinates, interspersed with stochastic transitions between
the energy levels of the subsystem.  Despite the similarities
to a recently introduced scheme based on a
Trotter decomposition of the quantum-classical propagator~\cite{trotter},
the basic version of the SSTP algorithm is not as stable at long times
and it also displays problems in the region of intermediate and strong
coupling to the bath, as illustrated by the results of calculations on
the spin-boson model~\cite{sergi-theor-chem,1st-energy-sampling-paper,2nd-energy-sampling-paper}.
The growth of the statistical error
in time can be mitigated by means of filtering
schemes. One such a scheme~\cite{brutalfilter} is essentially
based on establishing a cutting threshold of the observable when
it becomes too large because of the accumulation in time of the sampling weight.
Such a scheme will be referred to in the following as the observable-cutting scheme.
More recently, another filtering algorithm, which is 
based on a generalised sampling of non-adiabatic transitions,
has been proposed. Such an algorithm has been proven
to dramatically reduce the statistical error at 
long time~\cite{1st-energy-sampling-paper,2nd-energy-sampling-paper}.
This other scheme will be called in this paper as the transition-filtering scheme.

In this work, we use the SSTP algorithm to integrate the quantum-classical
Liouville equation for the spin-boson model
and perform a comparison of the performances of the two filtering schemes 
discussed above.
The main result of this paper is that the SSTP algorithm used in 
conjunction with the combination of 
the observable-cutting and the transition-filtering schemes
performs as well as the Trotter algorithm also in the intermediate and
strong coupling regimes.
This result is desirable since the SSTP algorithm is easier
to implement than its Trotter counterpart, especially when the
number of quantum states greater than two must be considered.
This promises to be advantageous when studying quantum systems
which are more complex than the spin-boson model.

The structure of the paper is as follows. Section~\ref{sec:qcle}
sketches the derivation of the quantum-classical
Liouville equation and its representation in the adiabatic basis.
In Sec.~\ref{sec:sstp} the basic version of the SSTP algorithm is illustrated
together with the observable-cutting and the transition-filtering schemes.
In the same section, the combined filtering scheme is introduced.
Section~\ref{sec:num} discusses the results of the numerical
calculations on the dynamics of the spin-boson model
using the various filtering schemes.
Finally, our conclusions are given in Sec.~\ref{sec:concl}.


\section{Quantum-classical Liouville equation}
\label{sec:qcle}

Let us consider a system that is defined by the following
Hamiltonian operator:
\begin{eqnarray}
\hat{H} = \hat{H}_{\rm S} + \hat{H}_{\rm B} + \hat{H}_{\rm SB}\,,
\label{eq:Hsys}
\end{eqnarray}
\noindent
where S, B and SB are subscripts denoting the subsystem, bath
and the coupling, respectively.
The Heisenberg equation of motion for an arbitrary operator
$\hat{A}$ can be written in symplectic form as~\cite{sergi-temp1}
\begin{eqnarray}
\frac{\partial}{\partial t}\hat{A} = \frac{i}{\hbar}\left[\begin{array}{cc}\hat{H}
& \hat{A}\end{array}\right]\cdot{\mathcal{B}}^c\cdot\left[\begin{array}{c}\hat{H}\\
\hat{A}\end{array}\right]\,,
\end{eqnarray}
where the symplectic matrix \cite{goldstein} ${\mathcal{B}}^c$ is given by
\begin{eqnarray}
{\mathcal{B}}^{c} = \left[\begin{array}{cc}0 & 1 \\ -1 & 0 \end{array}\right]\,.
\end{eqnarray}
It is assumed that the Hamiltonian of the bath depends on a pair of
canonically conjugate operators, $\hat{X} = (\hat{R}, \hat{P})$, and that the
coupling Hamiltonian $\hat{H}_{\rm SB}$ depends only on $\hat{R}$ and not $\hat{P}$.
The partial Wigner transform
of the operator $\hat{A}$ over the bath coordinates is
\begin{eqnarray}
\hat{A}_{\rm W}(X) = \int dz\, e^{i Pz/\hbar}\Big\langle R - \frac{z}{2}\Big|\hat{A}\Big|
R + \frac{z}{2}\Big\rangle\,.
\end{eqnarray}
The partial Wigner transform 
of the density matrix $\hat{\rho}$
of the system described by the Hamiltonian in Eq.~(\ref{eq:Hsys}) is
\begin{eqnarray}
\hat{\rho}_{\rm W}(X) = \frac{1}{\left(2\pi\hbar\right)^{3N}}\int dz\, e^{iPz/\hbar}
\Big\langle R - \frac{z}{2}\Big|\hat{\rho}\Big|R + \frac{z}{2}\Big\rangle\,,
\end{eqnarray}
where $X = (R,P)$ are now no longer operators but
canonically conjugate classical phase space variables.
The partial Wigner transform of the Heisenberg equation of motion
can be written in matrix form
upon introducing the antisymmetric matrix operator $\mathcal{D}$
given by~\cite{sergi-temp1}
\begin{eqnarray}
{\mathcal{D}} = \left[\begin{array}{cc} 0 & e^{\frac{i\hbar}{2}\stackrel{\leftarrow}
\partial_{k}{\mathcal{B}}^{c}_{kj}\stackrel{\rightarrow}\partial_{j}} \\ 
-e^{\frac{i\hbar}{2}\stackrel{\leftarrow}\partial_{k}{\mathcal{B}}^{c}_{kj}
\stackrel{\rightarrow}\partial_{j}} & 0 \end{array}\right]\,.
\end{eqnarray}
The symbols$\stackrel{\leftarrow}\partial_{k}=\stackrel{\leftarrow}\partial/\partial X_{k}$
and $\stackrel{\rightarrow}\partial_{k}=\stackrel{\rightarrow}\partial/\partial X_{k}$ 
denote the operators of derivation with respect to the phase-space coordinates 
acting to the left and right, respectively. 
The summation over repeated indices must be understood here and in the following.
The partial Wigner-transformed Hamiltonian can be written as
\begin{eqnarray}
\hat{H}_{\rm W}(X)  = \hat{H}_{\rm S} + H_{\rm W,B}(X) + \hat{H}_{\rm W,SB}(R)\,,
\label{eq:parW-H}
\end{eqnarray}
where we have assumed that the bath dependence of the coupling terms
is on positions only:
\begin{eqnarray}
\hat{H}_{\rm W,SB} = V_{\rm B}(R)\otimes\hat{H}'_{\rm S}\,,
\end{eqnarray}
where $H'_{S}$ acts only in the Hilbert space of the subsystem.
The above representation is equivalent to the Heisenberg representation, but in general
calculations are difficult to perform. However, in many instances a quantum-classical
approximation can be taken by means of a linear expansion of the
exponential terms in the $\mathcal{D}$ matrix, giving
\begin{eqnarray}
{\mathcal{D}}_{lin} = \left[\begin{array}{cc} 0 & 1 + \frac{i\hbar}{2}\stackrel{\leftarrow}\partial_{k}
{\mathcal{B}}^c_{kj}\stackrel{\rightarrow}\partial_{j} \\ -1 - \frac{i\hbar}{2}
\stackrel{\leftarrow}\partial_{k}{\mathcal{B}}^{c}_{kj}\stackrel{\rightarrow}\partial_{j} & 0
\end{array}\right]\,.
\end{eqnarray}
This allows one to write the quantum-classical Liouville equation as
\begin{eqnarray}
\frac{\partial}{\partial t}\hat{A}(X,t) 
&=&\frac{i}{\hbar}\left[\begin{array}{cc}
\hat{H}_{W}(X) & \hat{A}_{W}(X,t)\end{array}\right]\cdot{\mathcal{D}}_{lin}
\nonumber\\
&\cdot &
\left[\begin{array}{c}\hat{H}_{W}(X) \\ \hat{A}_{W}(X,t)\end{array}\right]\,.
\label{eq:WH-equation-of-motion}
\end{eqnarray}
When $V_{\rm B}(R)$ is linear in $R$ and $H_{\rm W,B}$ is quadratic in the
bath coordinates the linear expansion is
exact (${\mathcal{D}}_{lin} = {\mathcal{D}}$) and quantum-classical dynamics
is equivalent to full quantum dynamics. 



In order to perform calculations,
Eq.~(\ref{eq:WH-equation-of-motion}) must be represented in a basis.
The adiabatic basis leads naturally to a splitting of
non-adiabatic and adiabatic terms, which is ideal for surface-hopping
algorithms. The adiabatic basis is defined as the solution to the eigenvalue equation
\begin{eqnarray}
\hat{h}(R)|\alpha; R\rangle = E_{\alpha}(R)|\alpha; R\rangle\,.
\end{eqnarray}
In this basis the quantum-classical evolution takes the form
\begin{eqnarray}
A^{\alpha\alpha'}_{W}(X,t) = \sum_{\beta\beta'}\left(e^{it\mathcal{L}}\right)_{\alpha\alpha',
\beta\beta'}A_{W}^{\beta\beta'}(X)\,,
\label{eq:quant-class-evolution}
\end{eqnarray}
where the quantum-classical Liouville operator \cite{qcl6} is given by
\begin{eqnarray}
i{\mathcal{L}}_{\alpha\alpha',\beta\beta'} &=& \left(i\omega_{\alpha\alpha'} + iL_{\alpha\alpha'}\right)
\delta_{\alpha\beta}\delta_{\alpha'\beta'} + J^{\rm MJ}_{\alpha\alpha',\beta\beta'}
\nonumber
\\
&=& i{\mathcal{L}}^{0}_{\alpha\alpha'}\delta_{\alpha\beta}\delta_{\alpha'\beta'}
+ J^{\rm MJ}_{\alpha\alpha',\beta\beta'}\,,
\end{eqnarray}
where the $\rm MJ$ superscript denotes that we have used the momentum-jump approximation
\cite{momj,momj2}. The Bohr frequency is defined as
\begin{eqnarray}
\omega_{\alpha\alpha'}(R) = \frac{E_{\alpha}(R) - E_{\alpha'}(R)}{\hbar}\,,
\end{eqnarray}
and the classical-like Liouville operator for the bath degrees of freedom
is given by
\begin{eqnarray}
iL_{\alpha\alpha'} = \frac{P}{M}\cdot\partialR + \half(F^{\alpha}_{W} + F^{\alpha'}_{W})
\cdot\partialP\,,
\end{eqnarray}
where $F^{\alpha}_{W}$ and $F^{\alpha'}_{W}$ are the Hellman-Feynman forces for
adiabatic energy surface $E_{\alpha}$ and $E_{\alpha'}$ respectively.

The operator $J^{\rm MJ}_{\alpha\alpha',\beta\beta'}$
is known as the transition operator in the momentum-jump approximation~\cite{momj,momj2},
and is responsible
for the non-adiabatic transitions in the quantum subsystem and the accompanying changes in
the bath momentum. 
It is given by
\begin{eqnarray}
J^{\rm MJ}_{\aabb} = {\mathcal{T}}^{\rm MJ}_{\alpha\rightarrow\beta}\delta_{\alpha'\beta} +
{\mathcal{T}}^{*\rm MJ}_{\alpha'\rightarrow\beta'}\delta_{\alpha\beta}\,,
\end{eqnarray}
where
\begin{eqnarray}
{\mathcal{T}}^{\rm MJ}_{\alpha\rightarrow\beta} = \frac{P}{M}\cdot d_{\alpha\beta}(R)\exp\left[{\half\frac{\Delta
E_{\alpha\beta}(R)d_{\alpha\beta}(R)}{\frac{P}{M}\cdot d_{\alpha\beta}(R)}\cdot\partialP}\right]\,.
\nonumber\\
\end{eqnarray}
In the momentum-jump approximation the back-reaction on the bath
(i.e., the change to bath momenta)
accompanying a non-adiabatic transition can be calculated analytically. If we consider an
$\alpha\rightarrow\beta$ transition, the momentum-shift approximated $J$ operator,
$J^{\rm MJ}$, produces a shift in the bath momenta $P$. This shift is defined as
\begin{eqnarray}
P\rightarrow P' = P + \Delta^{\rm MJ}_{\alpha\beta}P\,,
\end{eqnarray}
where
\begin{eqnarray}
\Delta^{\rm MJ}_{\alpha\beta}P 
&=&-(P\cdot\hat{d}_{\alpha\beta})\hat{d}_{\ab}
\nonumber\\
& + & \hat{d}_{\ab}\sign
(P\cdot\hat{d}_{\ab})\sqrt{(P\cdot\hat{d}_{\ab})^{2} + M\Delta E_{\ab}}\,.
\nonumber\\
\label{eq:mom-shift}
\end{eqnarray}
\noindent
The symbol $\hat{d}_{\ab}$ is the unit vector associated with the coupling vector in the
multidimensional space of all the particle coordinates. Note that in the above, we have
assumed that all the masses are the same, however, it is a simple matter to extend this
to a system where the masses are different. If we expand the square root on the right-hand
side of equation (\ref{eq:mom-shift}), we obtain the approximated form for the
momentum shift rule
\begin{eqnarray}
\tilde\Delta^{\rm MJ}_{\ab}P = \half\frac{\Delta E_{\ab}(R)}{\frac{P}{M}\cdot\hat{d}_{\ab}}\hat{d}_{\ab}\,.
\label{eq:approx-mom-shift}
\end{eqnarray}
Note that while the exact momentum shift in equation (\ref{eq:mom-shift}) exactly
conserves the energy, the approximated form in equation (\ref{eq:approx-mom-shift}) does not.
In our calculations, only the exact form of the momentum shift was used.


\section{Filtering schemes for the SSTP algorithm}
\label{sec:sstp}

The SSTP algorithm is derived upon considering the evolution along a quantum-classical
trajectory given by the solution of Eq.~(\ref{eq:quant-class-evolution})
as a series of sequential small time steps $\tau$. 
Hence, the short-time expression of the quantum-classical propagator
$\left[\exp\left(i\tau{\mathcal{L}}^{\rm MJ}\right)\right]_{\aabb}$
is approximated to linear order in time as
\begin{eqnarray}
&&e^{i\tau{\mathcal{L}}^{0}_{\alpha\alpha'}}
\left(\delta_{\ab}\delta_{\alpha'\beta'} + \tau J^{\rm MJ}_{\aabb}\right)
\nonumber \\
&&= {\mathcal{W}}_{\alpha\alpha'}(\tau)e^{iL_{\alpha\alpha'}\tau}\left(\delta_{\ab}
\delta_{\alpha'\beta'} + \tau J^{\rm MJ}_{\aabb}\right)\,.
\label{eq:short-time-propagator}
\end{eqnarray} 
In the limit $\tau\to 0$, the concatenation of the short time steps
exactly reproduces the Dyson integral expansion of
the operator $\exp\left(i\tau{\mathcal{L}}\right)_{\aabb}$ for finite times~\cite{sstp}.
The computational evaluation of each single step $\tau$
can be evaluated upon considering the short time
propagator in Eq.~(\ref{eq:short-time-propagator}) as a stochastic operator.
The action of the
transition operator $J^{\rm MJ}_{\aabb}$ 
is then sampled using a suitable transition probability. This transition
probability is not uniquely fixed but has to be chosen following
the criteria of physical reasonability and computational efficiency.

The transition probability is defined as the probability of a non-adiabatic transition occurring
in a time interval $\tau$. A basic choice for this probability is given by
\begin{eqnarray}
{\mathcal{P}}^{0}_{\ab}(X,\tau) = \frac{\tau|\frac{P}{M}\cdot d_{\ab}(R)|}{1 + \tau|\frac{P}{M}\cdot
d_{\ab}(R)|}\,.
\end{eqnarray}
This transition probability then defines the probability of no transition occurring in the same
time interval as
\begin{eqnarray}
{\mathcal{Q}}^{0}_{\ab}(X,\tau) &=& 1 - {\mathcal{P}}^{0}_{\ab}
\nonumber
\\
&=& \frac{1}{1 + \tau|\frac{P}{M}\cdot d_{\ab}(R)|}\,. 
\end{eqnarray} 
when at time step $i$ in the calculation the transition probability is sampled, and a transition
occurs, the observable is multiplied by a factor of 
$\left(P/M\right)\cdot d_{\ab}\left({\mathcal{P}}^{0}_{\ab}\right)^{-1}$.
If no transition occurs, then the observable is multiplied by a factor
of $\left({\mathcal{Q}_{\ab}}^{0}\right)^{-1}$.
The $\left(P/M\right)\cdot d_{\ab}$ term introduced when a transition
occurs originates from the action
of the $J^{\rm MJ}_{\aabb}$ operator.
The concatenation of these factors result in a weight in the observable which causes
an error in the result at longer times. We thus need to sample non-adiabatic transitions in 
such a way that minimises this statistical error.



The first method for reducing statistical error tackles the problem directly. It is a simple but effective
approach. As mentioned above, the magnitude of the weight, which is used to calculate the observable,
grows with time and causes the value of the observable to grow, leading to large statistical error. Knowing
this, we can introduce a threshold value $c_{t}$ which sets an upper bound to the magnitude of the weight.
If at a stage $j$ in the calculation of a trajectory we have that the magnitude of the weight $W$ becomes larger
than $c_{t}$, it is instead set to the value of $c_{t}$. Mathematically, we can write this as:
\begin{eqnarray}
W =  \left\{\begin{array}{cc} W & \mbox{if $|W| < c_{t}$} \\ \sign(W)c_{t} & \mbox{if $|W| > c_{t}$}\end{array}\right.
\end{eqnarray}
Note that this cutting only affects the magnitude of the weight, the sign remains the same. This cutting
ensures that the weight can never grow to values where a single trajectory is having an overly large effect
on the value of the observable. Consequently, we do not see the large statistical error in the result
at longer times. While effective, however, this scheme does not have any 
physical basis, unlike the transition filtering scheme.


Another approach to the reduction of the statistical error has been
recently proposed in Refs.~\cite{1st-energy-sampling-paper,2nd-energy-sampling-paper}.
Essentially, It is based on filtering out those non-adiabatic transitions which would lead 
to too big a change in the momenta.
In order to illustrate such an approach, it is useful
to recall the form of the energy variation because of a non-adiabatic transition
calculated using an approximate form of the momentum-shift rule:
\begin{eqnarray}
{\mathcal{E}}_{\ab} = \frac{P'^{2}}{2M} + E_{\alpha}(R) - \left(\frac{P^2}{2M} + E_{\beta}(R)\right)\,,
\end{eqnarray}
\noindent
where $P' = P + \tilde\Delta^{\rm MJ}_{\alpha\beta}P$.
Upon introducing the parameter $c_{\mathcal{E}}$
and the weight $\omega(c_{\mathcal{E}},{\mathcal{E}}_{\ab})$,
one can define a generalised transition probability
\begin{eqnarray}
{\mathcal{P}}^{EC}_{\ab}(X,\tau) = \frac{\tau|\frac{P}{M}\cdot d_{\ab}(R)|\omega(c_{\mathcal{E}},{\mathcal{E}_{\ab}})}
{1 + \tau|\frac{P}{M}\cdot d_{\ab}(R)|\omega(c_{\mathcal{E}},{\mathcal{E}_{\ab}})}\,.
\label{eq:prob-gen}
\end{eqnarray}
\noindent
This in turn defines the probability of no transition occurring as
\begin{eqnarray}
{\mathcal{Q}}_{\ab}(X,\tau) &=& 1 - {\mathcal{P}}^{EC}_{\ab}
\nonumber
\\
&=& \frac{1}{1 + \tau|\frac{P}{M}\cdot d_{\ab}(R)|\omega(c_{\mathcal{E}},{\mathcal{E}_{\ab}})}\,.
\label{eq:prob-no-gen}
\end{eqnarray}
The weight $\omega(c_{\mathcal{E}},{\mathcal{E}_{\ab}})$ is defined as
\begin{eqnarray}
\omega(c_{\mathcal{E}},{\mathcal{E}_{\ab}}) = \bigg\lbrace\begin{array}{cc} 1 & \mbox{if ${\mathcal{E}}_{ab}\leq
c_{\mathcal{E}},$} \\ 0 & \mbox{otherwise.}\end{array}
\end{eqnarray}
The transition probabilities in Eqs.~(\ref{eq:prob-gen}) and~(\ref{eq:prob-no-gen})
allow one to control the amplitude of energy fluctuations
that would be caused by an approximate momentum shift
through the use of the numerical parameter $c_{\mathcal{E}}$.
Whenever a non-adiabatic transition
would cause a virtual energy fluctuation that is larger than $c_{\mathcal{E}}$,
the transition probability
becomes zero, and no transition can occur.
This generalisation of the basic sampling scheme
allows non-adiabatic transitions to occur only in regions
where the approximate momentum shift rule causes small virtual variation
of the energy of the system: this happens when the change in the momentum
is not too big.
Such a scheme has been proven numerically to be very efficient in
reducing statistical error at 
long times~\cite{1st-energy-sampling-paper,2nd-energy-sampling-paper}.


Since each of the above filtering techniques approach
the statistical error problem from different
angles, it is interesting to combine them within a single simulation algorithms.
According to such an idea, in each simulation,
the non-adiabatic transitions are filtered according to the 
transition-filtering scheme, using the parameter $c_{\mathcal{E}}$, 
in addition to the observable being
cut when it grows too large, according to the parameter $c_{t}$.


\section{Numerical calculations}
\label{sec:num}

Our numerical study was performed on the spin-boson model~\cite{leggett},
which can be considered as a paradigmatic example for quantum dynamics~\cite{sergi-theor-chem}
for which the adiabatic states are known exactly.
Such a system comprises a single spin coupled to a bath of harmonic oscillators.
Using adimensional coordinates~\cite{sstp,trotter,sergi-theor-chem},
the spin-boson Hamiltonian is given by
defining the various terms in Eq.~(\ref{eq:parW-H}) as
\begin{eqnarray}
\hat{H}_{\rm S}&=& -\Omega\hat{\sigma}_{x}\;, \\
H_{\rm W,B}&=& \sum_{i=1}^{N}\left(\frac{P_{i}^2}{2M_{i}} + \half M_{i}\omega_{i}^{2}R_{i}^{2}\right) \;,\\
\hat{H}_{\rm W,SB}&=&- \sum_{i=1}^{N}c_{i}R_{i}\hat{\sigma}_{z}\,,
\end{eqnarray}
where $\hat{\sigma}_{x}$ and $\hat{\sigma}_{z}$ are the Pauli spin matrices and $c_{i}$
are the coupling coefficients. These coefficients are determined by requiring that
the system spectral density is Ohmic~\cite{makarov}.

In simulations, we have set the spin in an excited state at $t=0$, and the quantum harmonic
modes are at thermal equilibrium, with no coupling before $t=0$. After $t=0$, the coupling
is switched on, and we calculate the observable $\langle\hat{\sigma}_{z}(t)\rangle$,
or population difference of the system.
We have considered $n=2$ non-adiabatic transitions per trajectory,
as this was sufficient for the results to converge.
Each simulation used a total of $N_{\rm mcs} = 10^5$
sampled phase space points for the initial conditions.
The integration time step was $dt = 0.1$ in dimensionless units.

\begin{figure}
\includegraphics[width=1.\linewidth]{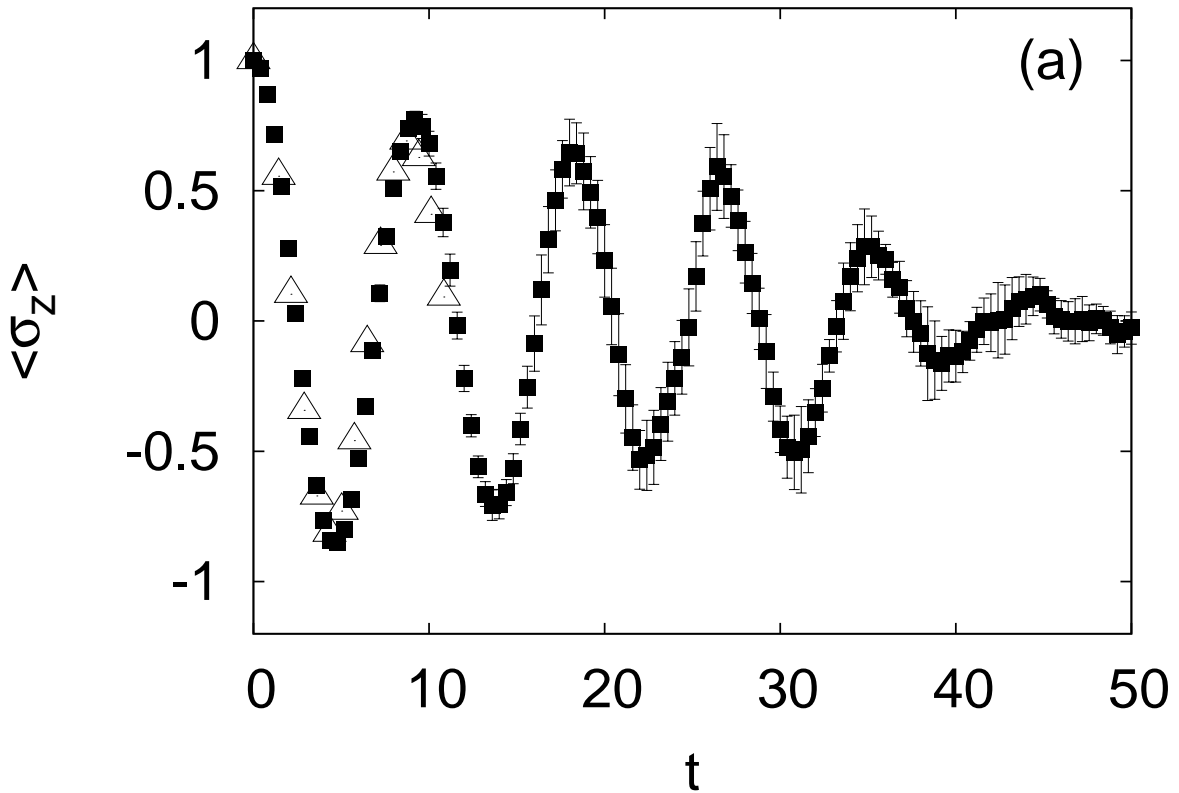}
\caption{Comparison of the 
SSTP results with the observable-cutting ($\blacksquare$)
and the transition-filtering algorithm ($\blacklozenge$, panel b)
to exact quantum results $(\triangle)$.
System parameters were $\beta = 0.3$, $\xi = 0.007$, $\Omega = 1/3$,
corresponding to weak coupling. The value of the
threshold parameter for the observable-cutting was $c_{t} = 100.0$.
Two non-adiabatic
transitions were included in the calculations.}
\label{fig:fig1}
\end{figure}

\begin{figure}
\includegraphics[width=1.\linewidth]{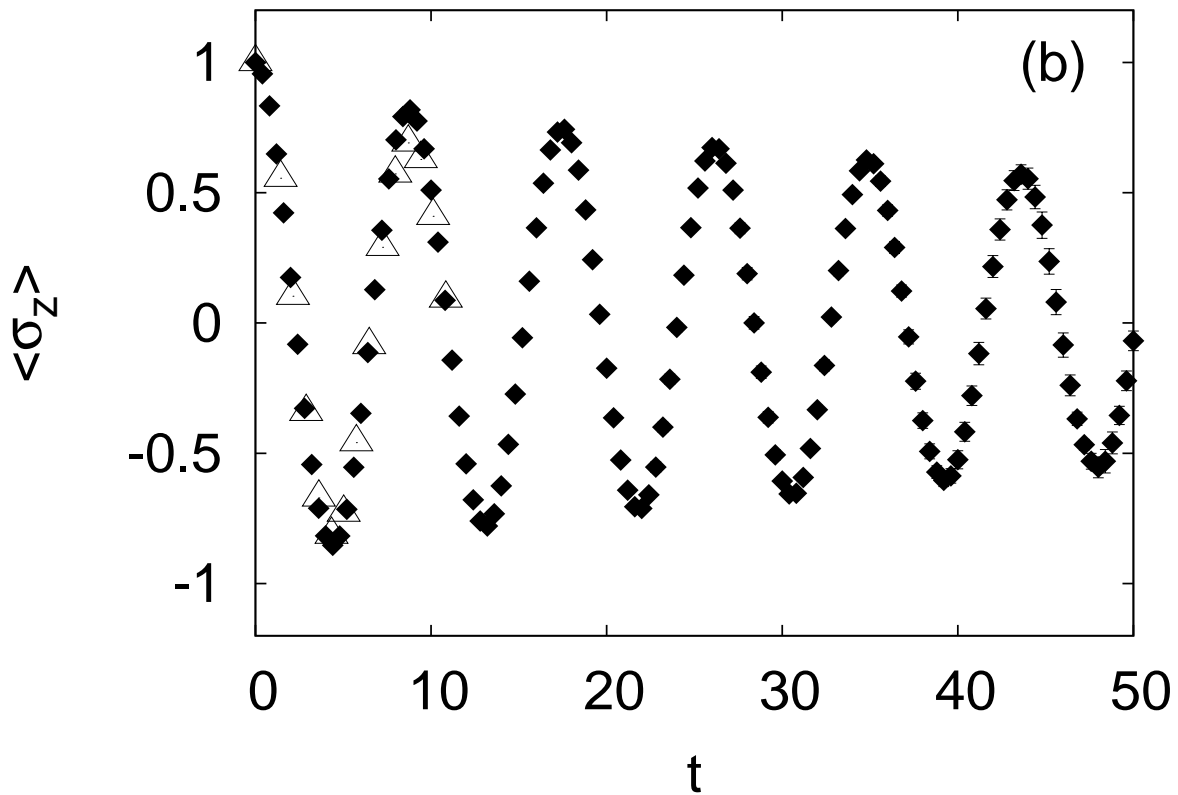}
\caption{Comparison of the transition-filtering results ($\blacklozenge$)
to exact quantum results $(\triangle)$.
System parameters were $\beta = 0.3$, $\xi = 0.007$, $\Omega = 1/3$,
corresponding to weak coupling. The value of the
threshold parameter for the observable-cutting was $c_{t} = 100.0$, and the value of the control
parameter for the energy conserving filtering was $c_{\mathcal{E}} = 0.005$. Two non-adiabatic
transitions were included in the calculations.}
\label{fig:fig2}
\end{figure}

\begin{figure}
\includegraphics[width=1.\linewidth]{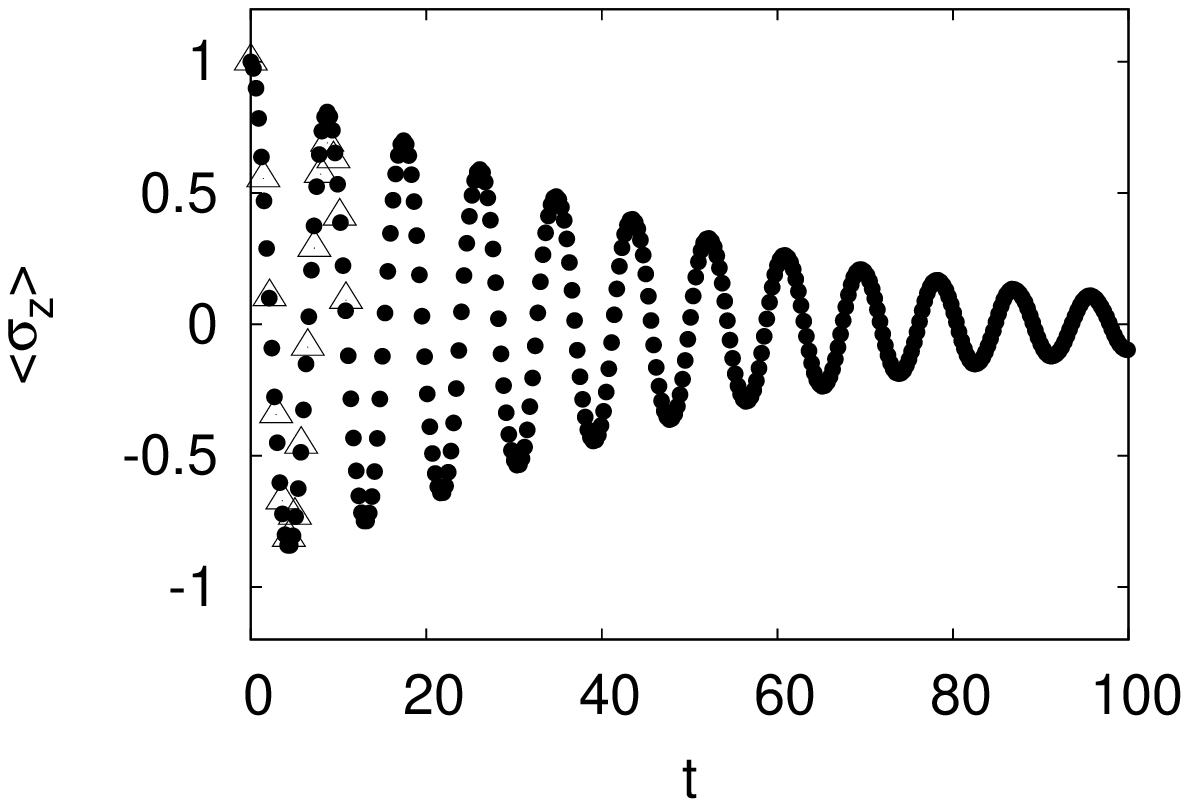}
\caption{Comparison of
the SSTP with the combined filtering algorithm $(\bullet)$
to exact quantum results $(\triangle)$.
System parameters were $\beta = 0.3$, $\xi = 0.007$, $\Omega = 1/3$.
The value of the threshold parameter was $c_{t} = 1.5$, and the value of the energy
conserving filtering control parameter was $c_{\mathcal{E}} = 0.005$. Two non-adiabatic
transitions were included in the calculations.}
\label{fig:fig3}
\end{figure}

Figures~\ref{fig:fig1} and~\ref{fig:fig2}
give the results for weak coupling, with system parameters $\beta = 0.3$,
$\xi = 0.007$, and $\Omega = 1/3$. From Figs.~\ref{fig:fig1} and ~\ref{fig:fig2}, we see that both
filtering techniques
give results that agree well with the influence functional 
path integral calculations~\cite{makarov}, however, the
two results deviate from each other at longer times. For both cases we can observe
the growth of the statistical
error at longer times, although it is relatively minimal for weak coupling. Figure~\ref{fig:fig3} 
displays the weak coupling result for the combined filtering scheme.
In this case we again see the
excellent agreement with the influence functional results, but the error bars are smaller
than the points for the entire simulation time. 
Moreover, the calculation remains stable for 
longer times~\cite{1st-energy-sampling-paper,2nd-energy-sampling-paper}
than those obtained in previously published results.

\begin{figure}
\includegraphics[width=1.\linewidth]{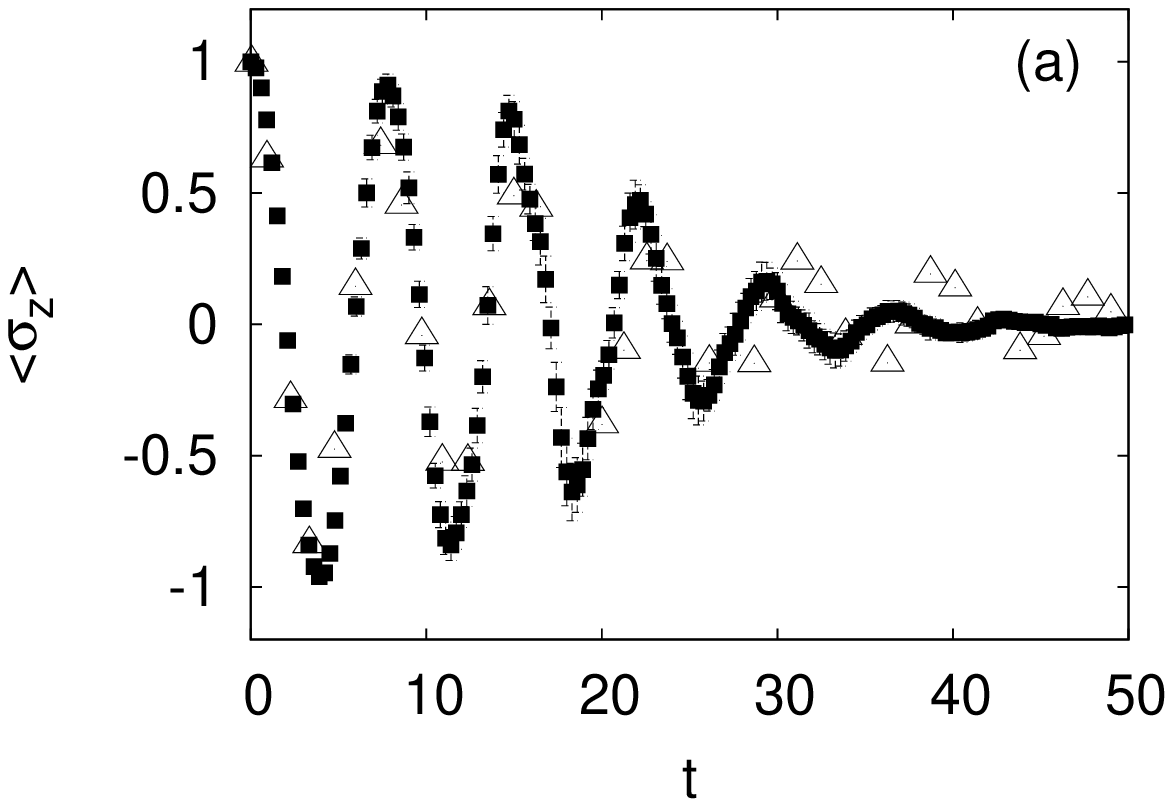}
\caption{Comparison of the 
SSTP with the observable-cutting ($\blacksquare$)
to exact quantum results $(\triangle)$. 
System parameters were $\beta = 12.5$, $\xi = 0.09$, $\Omega = 0.4$, corresponding to 
mid-range coupling. The value
of the threshold parameter for the direct filtering was $c_{t} = 50.0$.
Two
non-adiabatic transitions were included in the calculations.}
\label{fig:fig4}
\end{figure}

\begin{figure}
\includegraphics[width=1.\linewidth]{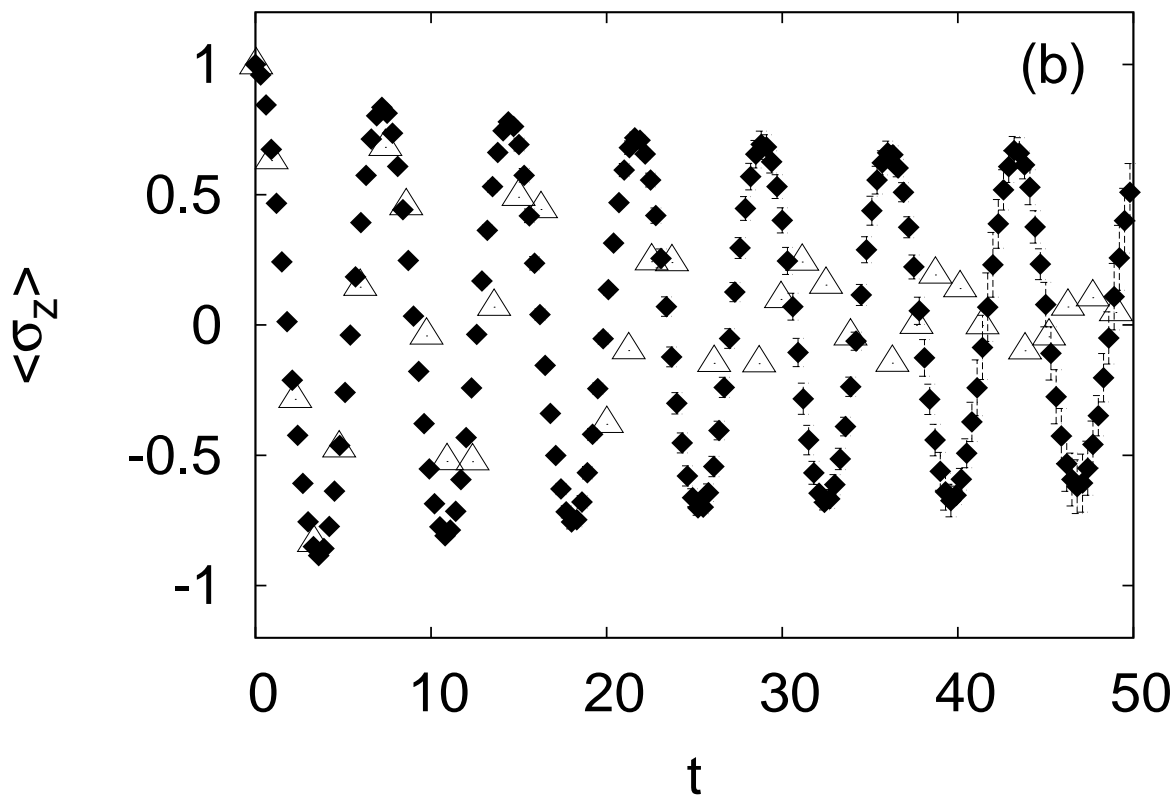}
\caption{Comparison of the 
transition-filtering algorithm ($\blacklozenge$)
to exact quantum results $(\triangle)$. 
System parameters were $\beta = 12.5$, $\xi = 0.09$, $\Omega = 0.4$, corresponding to 
mid-range coupling. The value
of the control parameter was $c_{\mathcal{E}} = 0.025$. Two
non-adiabatic transitions were included in the calculations.}
\label{fig:fig5}
\end{figure}

\begin{figure}
\includegraphics[width=1.\linewidth]{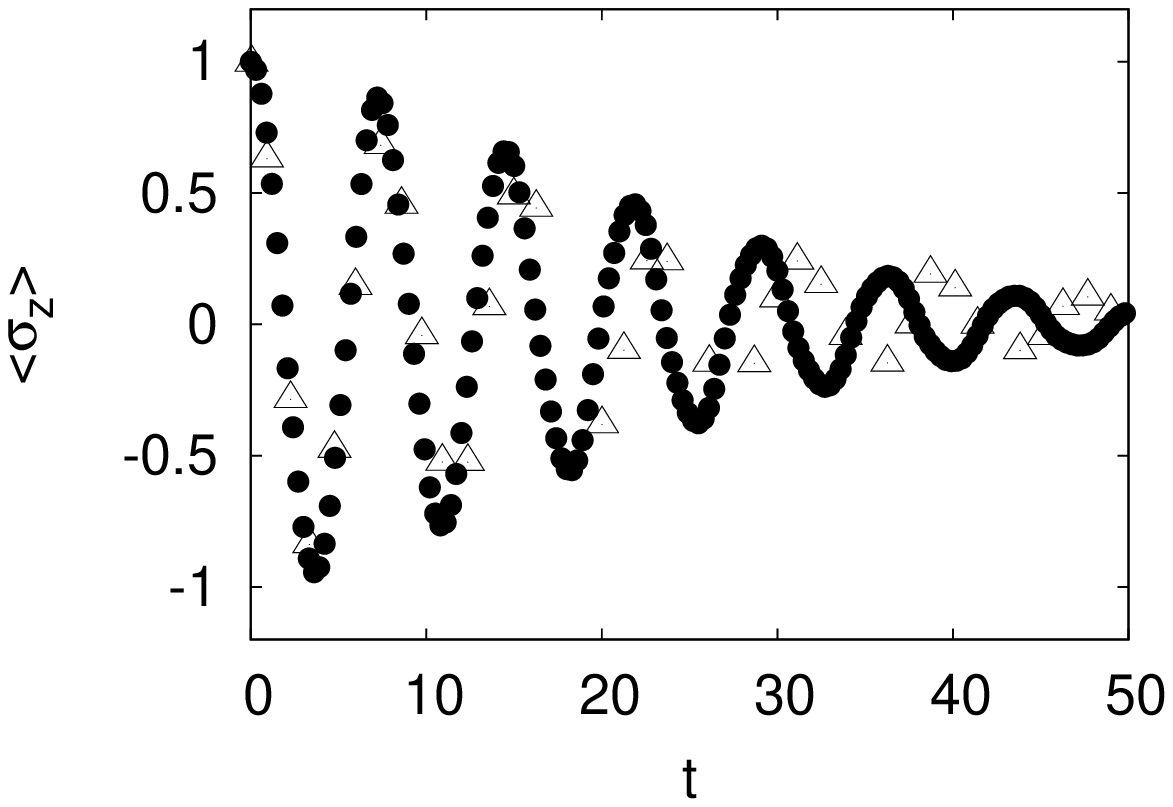}
\caption{Comparison of the 
SSTP with the combined filtering algorithm $(\bullet)$
to exact quantum results $(\triangle)$.
System parameters were $\beta = 12.5$, $\xi = 0.09$, $\Omega = 0.4$.
The value of the threshold parameter was $c_{t} = 3.5$, and the value of the 
control parameter was $c_{\mathcal{E}} = 0.05$. Two non-adiabatic
transitions were included in the calculations.}
\label{fig:fig6}
\end{figure}

Figures~\ref{fig:fig4} and~\ref{fig:fig5} show the results
for mid-range coupling. The system parameters used were $\beta = 12.5$, $\xi = 0.09$ and
$\Omega = 0.4$.  Figures~\ref{fig:fig4} and~~\ref{fig:fig5} give
the comparison of the two filtering schemes. 
The results for both
filtering schemes agree very well with the exact quantum result from Ref.~\cite{makarov} 
up until approximately $t = 20$, but after this time the results deviate somewhat.
In the case of the transition-sampling filter,
we do not observe the damping that occurs in the exact result - the oscillations
remain large. For the observable-cutting scheme, however, we see the opposite. 
The observable-cutting filter damps
the result too much at longer times, causing it to become zero. In 
Fig.~\ref{fig:fig6},
we have the result
for the combined filtering scheme.
We see a dramatic improvement over both the individual filtering
scheme, since the combined filter does not exhibit either of the problems observed above.
The combined filtering result agrees far better with the exact quantum result at longer times, 
with the error bars smaller than the points for the entire simulation time. 

\begin{figure}
\includegraphics[width=1.\linewidth]{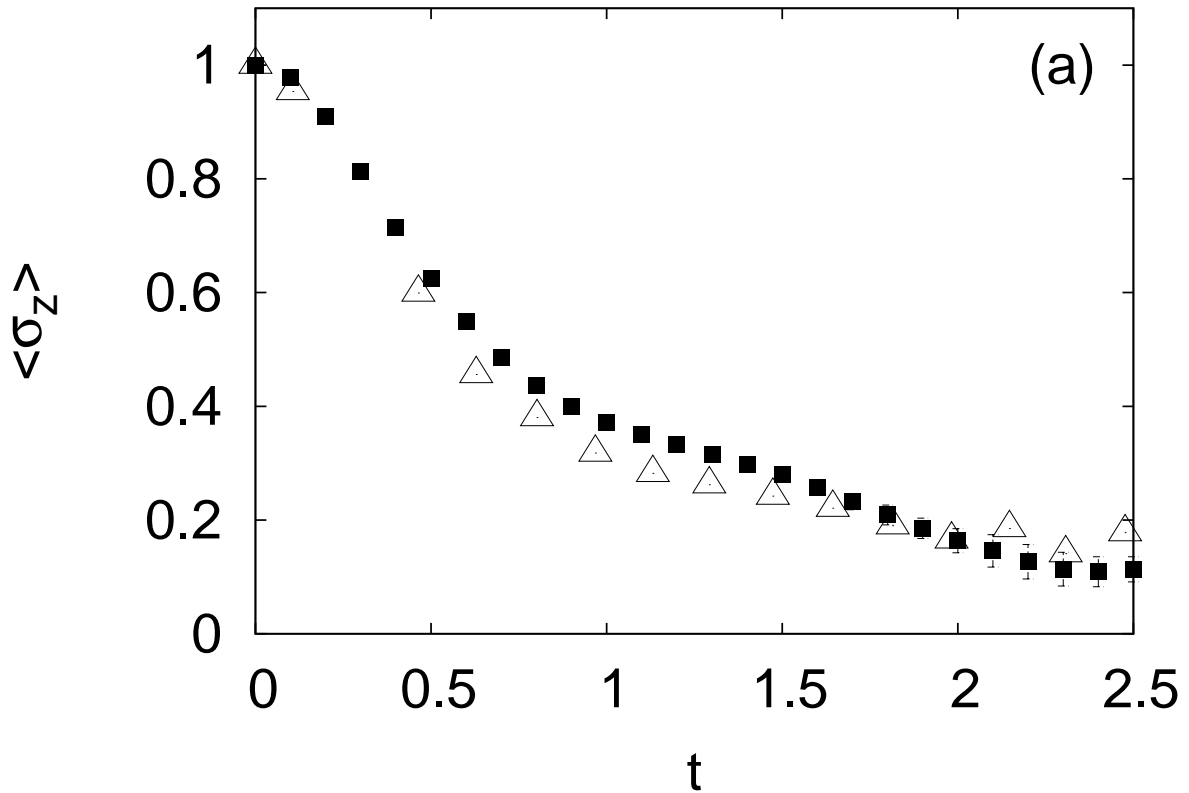}
\caption{Comparison of the
SSTP with the observable-cutting ($\blacksquare$)
to exact quantum results $(\triangle)$. 
System parameters
were $\beta = 0.25$, $\xi = 2.0$, $\Omega = 1.2$, corresponding to high coupling.
The value of the
bound parameter for the direct filtering was $c_{t} = 50.0$.
Two non-adiabatic transitions were
included in the calculations.}
\label{fig:fig7}
\end{figure}

\begin{figure}
\includegraphics[width=1.\linewidth]{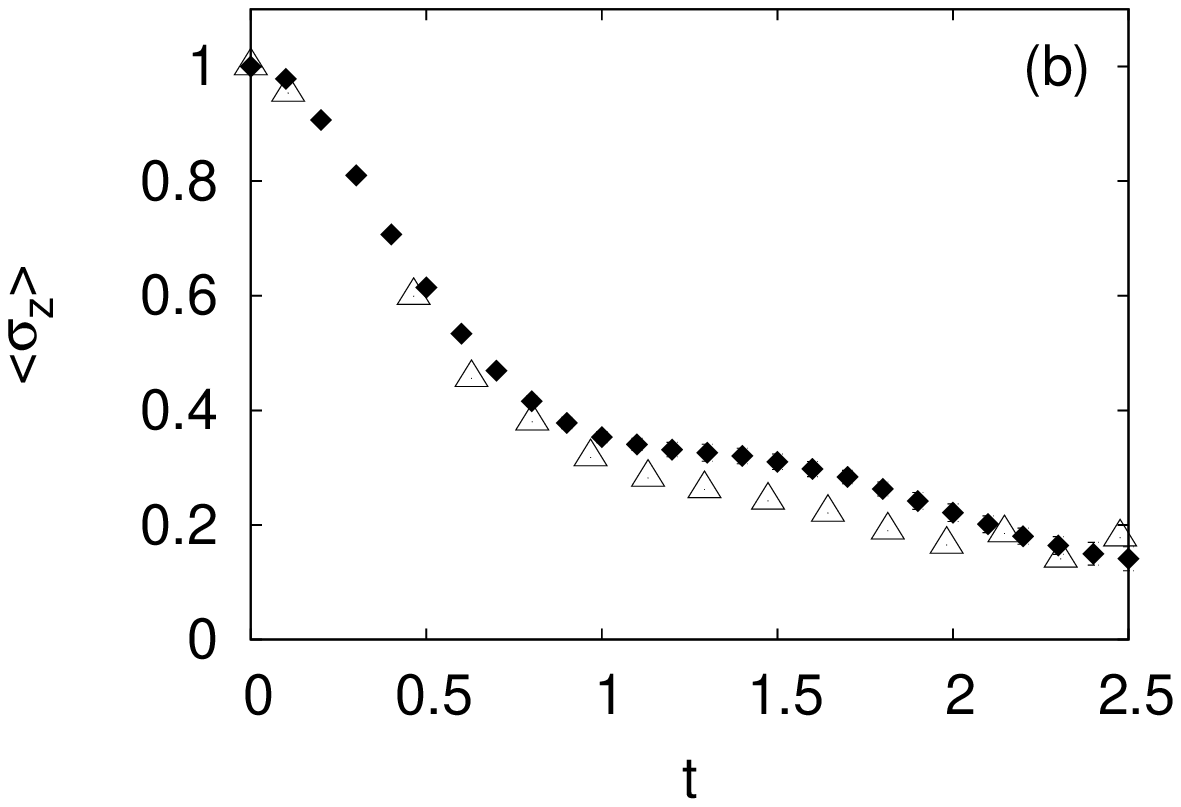}
\caption{Comparison of the
transition-filtering algorithm ($\blacklozenge$, panel b)
to exact quantum results $(\triangle)$. 
System parameters
were $\beta = 0.25$, $\xi = 2.0$, $\Omega = 1.2$, corresponding to high coupling.
The value of the control parameter
was $c_{\mathcal{E}} = 0.5$. Two non-adiabatic transitions were
included in the calculations.}
\label{fig:fig8}
\end{figure}

\begin{figure}
\includegraphics[width=1.\linewidth]{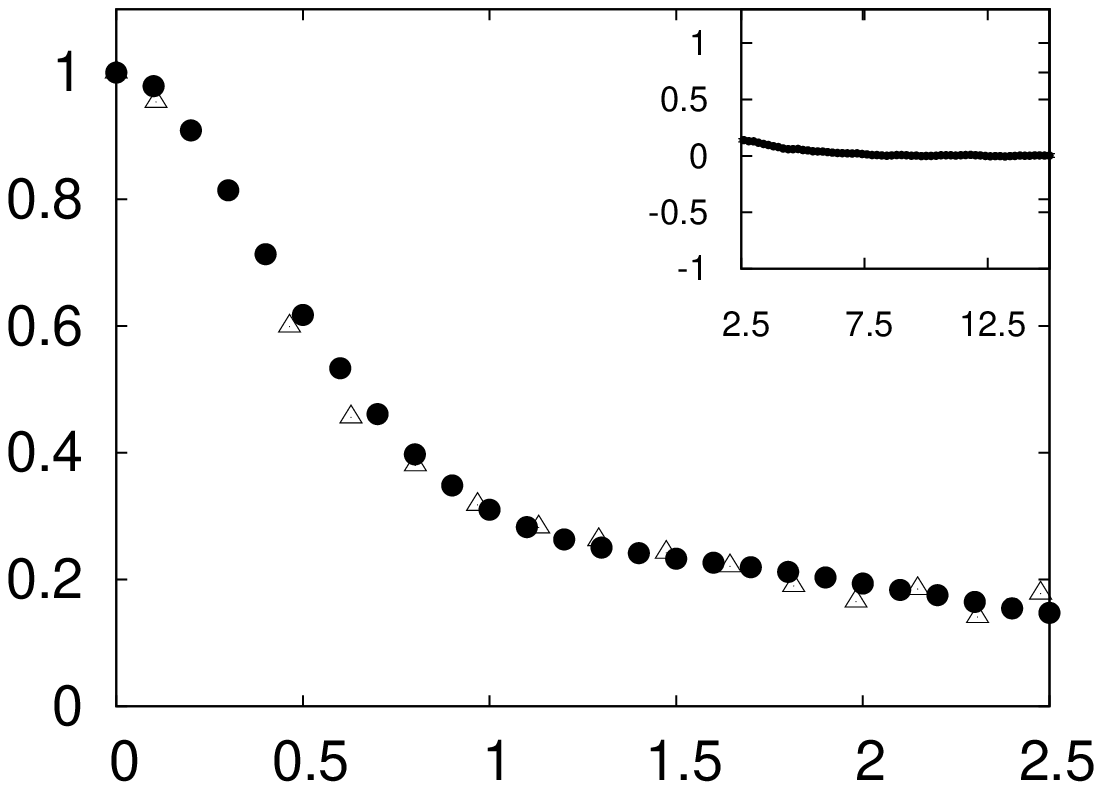}
\caption{Comparison of the 
SSTP with the combined filtering algorithm $(\bullet)$
to exact quantum results $(\triangle)$.
System parameters were $\beta = 0.25$, $\xi = 2.0$, $\Omega = 1.2$.
The value of the threshold parameter was $c_{t} = 5.0$, and the value of the 
control parameter was $c_{\mathcal{E}} = 1.0$. Two non-adiabatic
transitions were included in the calculations.}
\label{fig:fig9}
\end{figure}

In Figs.~\ref{fig:fig7} and~\ref{fig:fig8} the results for strong coupling are presented. 
For these results we adopted the system
parameters $\beta = 0.25$, $\xi = 2.0$ and $\Omega = 1.2$. 
From Fiasg.~\ref{fig:fig7} and~\ref{fig:fig8}
we can see that
the two filtering schemes are incapable of reproducing the exact quantum results of 
Ref.~\cite{golosov}
at even short times. Although both schemes are successful at reducing the statistical error, 
we do see that the error bars become larger than the points at approximately $t = 2$.
In Fig.~\ref{fig:fig9} we show the result obtained with the combined filtering scheme
for strong coupling. Again, the improvement is remarkable. In
the main figure, an excellent agreement with the exact quantum result is illustrated,
while the inset shows that the result can be extended to long times with 
statistical error remaining smaller than the points.

\section{Conclusions}
\label{sec:concl}

We have studied three different methods for reducing the statistical error
when simulating the quantum-classical Liouville approach to non-adiabatic dynamics
by means of the Sequential Short-Time Step algorithm~\cite{sstp}.
The first two methods are the observable-cutting scheme
(which uses the reset to threshold value for the statistical weights
entering the definition of the observable)
and the transition-filtering approach (which prunes the ensemble
of allowed non-adiabatic transitions on the basis of a generalised
sampling probability).
We have used the spin-boson model as a paradigmatic example
of quantum dynamics in a dissipative environment~\cite{leggett}
and performed numerical calculations on the evolution in time
of the state population difference of this model.
The use of either scheme gives rise to
results that have smaller statistical error than those obtained when using the basic
sampling, and both filtering techniques are capable of producing results in good
agreement with the numerically exact quantum results for short times, but only
for the intermediate and weak coupling regimes. Although both schemes are an
improvement over the basic sampling method in the SSTP algorithm,
they are still unable to reproduce the numerically
exact results for strong coupling, and fail at longer times for intermediate coupling
as well. Moreover, both schemes, when used in separation from the other,
are not able to curb the increase of the statistical error at longer times.

Nevertheless, we have shown that the combination of these two filtering methods
in a single scheme solve both of 
the problems encountered by the individual filtering schemes.
This is the main result of this paper.
Upon using the combined filtering scheme, we have produced results
that not only have negligible statistical error for longer simulation time
than that accessible in previously published calculations,
but compare far more favorably with the numerically exact results.
The combined method is able
to nearly perfectly reproduce the strong coupling results, whereas the individual schemes
could not do this even at very short times.
Our results are also as good as those obtained with the Trotter-based algorithm
for the simulation of the quantum-classical Liouville equation~\cite{trotter}.
However, since the SSTP algorithm is easier to implement for systems which have
a number of quantum states greater than two, our proposal of the combined filtering
scheme promises to be advantageous for more complex numerical studies of non-adiabatic 
dynamics.

\section*{Acknowledgements}
This work is based upon research supported by
the South African Research Chair Initiative of 
the Department of Science and Technology and
the National Research Foundation.
 

\end{document}